\documentclass[%
 reprint,
superscriptaddress,
 amsmath,amssymb,
 aps,
prb,
]{revtex4-2}

\usepackage{graphicx}
\usepackage{dcolumn}
\usepackage{bm}

\begin{document}

\title{The Low-Temperature Electronic Specific Heats of Disordered Ag-Au Alloys, Revisited}

\author{David Hinojosa-Romero}%
\author{Renela M. Valladares}%
\author{Alexander Valladares}%
	\affiliation{%
 	Facultad de Ciencias, Universidad Nacional Autónoma de México, Apartado Postal 70-542, Ciudad Universitaria, CDMX, 04510, México.
 	}%
\author{Isaías Rodríguez}%
\author{Ariel A. Valladares}%
 \email{valladar@unam.mx}%
	\affiliation{%
 	Instituto de Investigaciones en Materiales, Universidad Nacional Autónoma de México, Apartado Postal 70-360, Ciudad Universitaria, CDMX, 04510, México.
 	}%

\date{\today}

\begin{abstract}
Disordered alloys of silver and gold have been in the interest of the condensed matter community for decades since they are the prototype of the ideal solid solution due to the chemical similarity of their constituents and due to their potential industrial applications. Although they are considered well-known materials, surprises have appeared that have not been well understood despite several studies performed. One example are the experimental results of the electronic specific heat at low temperatures of disordered Ag-Au alloys. In 1966, Green and Valladares [Phys. Rev. \textbf{142}, 379 (1966)] conducted experimental studies of $\gamma$, the coefficient of the temperature in the expression for the electronic specific heat at low temperatures, finding a parabolic behavior as a function of the concentration, when a linear interpolation between the pure-element values was expected. This detonated several ulterior experiments that corroborated this parabolic behavior, and theoretical attempts followed that did not satisfactorily succeed at the explanation. It is our hope that this paper will contribute to the understanding of the experimental results; old problems can be reanalyzed with the help of new tools.
\end{abstract}

\maketitle

\section{\label{sec:INTRO}INTRODUCTION}

\begin{figure}
		\includegraphics{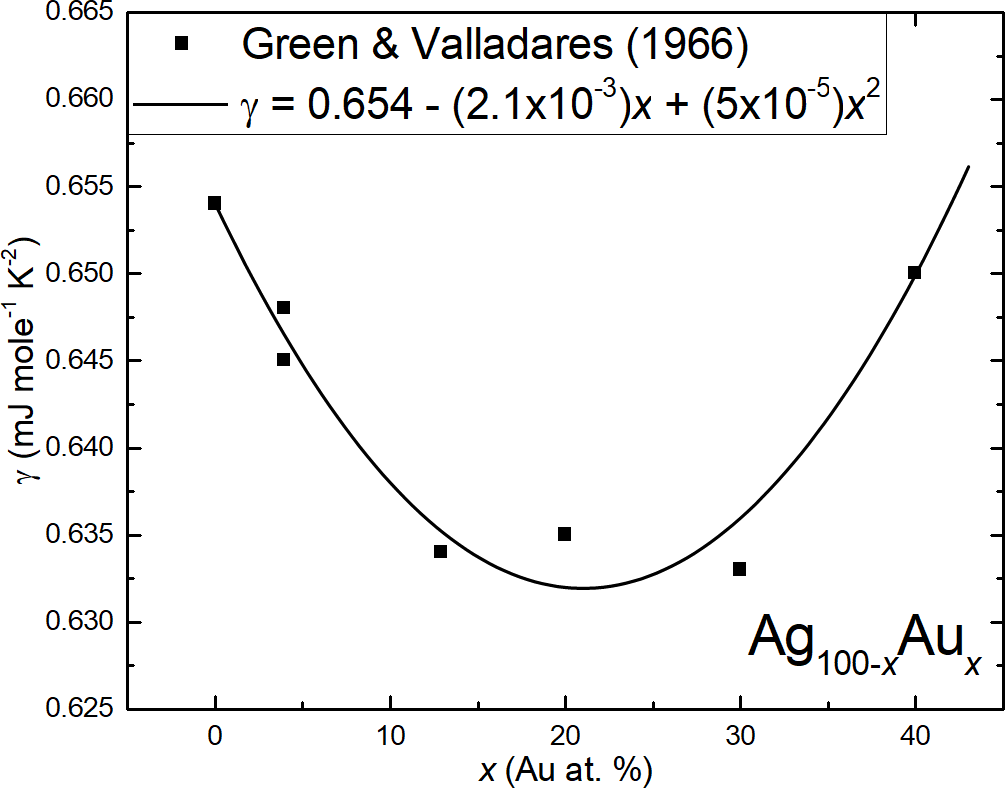}
		\caption{\label{fig:1} Pioneering measurements of the electronic specific heat coefficient $\gamma$ for the silver-rich Ag-Au alloys as reported by Green and Valladares \citep{Green1966}.}%
	\end{figure}

Almost 60 years ago, Green and Valladares \cite{Valladares1965, *Green1966} measured the low-temperature specific heats of silver-gold alloys as a function of gold concentration up to 40\% gold at liquid helium temperatures. Silver (Ag) and gold (Au) are so similar that the foretold result was a linear interpolation of the values for the two pure metals, and therefore nothing of interest was in it. However, the results revealed a non-linear (parabolic) behavior of $\gamma$, Figure \ref{fig:1}, which prompted new measurements and, on the theoretical side, new hypotheses that gave rise to new publications on the subject. At that time, the presence of iron impurities in the samples was suggested as the cause of this deviation, but the samples used in the experiment were 6 nines (99.9999\%) pure, and subsequent results reproduced this parabolic behavior as well. Theoretically, several mechanisms were proposed to try to understand this behavior, some based on perturbation theory, some invoking the contribution of other interactions, like the electron-phonon, to the specific heat, some using periodic crystalline structures although the alloys are in principle disordered materials. Nonetheless, none of these approaches were satisfactory and the approximations in the calculations obscured the essence of the results.

The first attempt to an explanation was made by Stern \cite{Stern1966} who invoked a charging effect around the solute element in resonance with the ideas propounded by Friedel's group for dilute alloys \cite{Friedel1954} at the time. In Stern's work, it was suggested that this charging effect made the energy of the electrons deviate from a linear interpolation between the values for the pure elements. Stern proposed that for Ag-Au alloys the properties near the Fermi surface could be treated by perturbation theory, a highly restrictive conclusion that would make it difficult to treat other binary systems since, as Stern argues, perturbation theory is valid when the charging effect is small. Therefore, a perturbation approach is not applicable to systems where the charging effect is prominent.

Then the electron-phonon interaction was considered to be the reason for the unexpected results, and calculations were reported that did not quite agree with the experiments \cite{Haga1967}. The renormalization of the electron mass due to the electron-phonon interaction has been cited as the reason for the discrepancies between experimental and theoretical results in transport measurements. Therefore it was natural to assume that it could also account for the parabolic behavior of the coefficient $\gamma$ in the electronic specific heat at low temperatures for these alloys. 

Finally, in 1989, a Linear Muffin-Tin Orbital (LMTO) calculation was reported that provides results for the electronic structure of the Ag-Au alloys displaying a chaotic behavior for the seven concentrations studied \cite{Kokko1989}. Real Ag-Au alloys are disordered structures but these authors used crystalline, ordered structures of the types AB$_{3}$ (simple cubic), AB$_{2}$ (body-centered tetragonal), and AB (simple tetragonal), to substitute solute atoms for solvent atoms. They believe that, in this manner, it is possible to explain the observed behavior of the density of electronic states at the Fermi level, ``at least qualitatively” \cite{Kokko1989}.

Some attempts to deal with this behavior within the Coherent Potential Approximation, (CPA), were carried out by Clark and Dower \cite{Clark1972} and they found an interesting behavior reminiscent of the experimental results. In all these approaches, crystallinity and Bloch states are used, which indicates the subjacent necessity to consider the disordered structure as an ordered one, as a first approximation. 

The experimental results of Martin in 1968 \cite{Martin1968}, and of Davis and Rayne in 1972 \cite{Davis1972}, which will be used extensively in this work, corroborated the parabolic tendency and extended the study to all concentrations of the solute, providing a more complete work than the one reported in 1966. These results were puzzling and awaited a better explanation. Clearly, perturbation theory was questioned as to whether it adequately described the phenomenon, and electron-phonon interaction was considered too weak to generate such a deviation from linearity. Furthermore, using crystallinity for the disordered structures always left an unsatisfying feeling of an inadequate approach, even as a first approximation.

For these reasons, we undertook a quantum mechanical study of these disordered supercells as a function of the positions of solvent and solute atoms, without any approximation, except for the random substitution of solvent atoms. In this manner we were able to determine the influence of the atomic arrangements on the electronic density of states in general, and its value at the Fermi surface in particular. 

The quantum mechanical tools we used are part of the Materials Studio Suite of codes and are based on Density Functional Theory (DFT) an approach well suited to describe the electronic properties of materials.  We aimed to obtain a description independent of the theoretical approach and perhaps only dependent on the concentration and specific arrangements of the atoms in a given supercell. The results found indicate that the parabolic behavior is due to the atomic structure of the supercell and that the electron-phonon mass renormalization contributes to the excellent agreement of our results with experiment.

This paper contributes to elucidate the results 60 years later, demonstrating the importance of re-studying unexplained old results with modern and more sophisticated tools.

\section{\label{sec:Preamble}PREAMBLE}

As is well known, silver and gold are neighbors in Group 11 of the periodic table with an electronic configuration of: (Kr)4d$^{10}$5s$^{1}$ for $^{47}$Ag; whereas $^{79}$Au, placed on the next row, has an electronic configuration of (Xe)4f$^{14}$5d$^{10}$6s$^{1}$. Their electronic structures are similar in the sense that both have filled d-orbitals and a single s-electron in the outer shell. Aditionally, the topological structures in the condensed state, are of the same type, i.e. face-centered cubic (FCC), with comparable lattice parameters: 4.0857 \AA\ for silver and  4.0783 \AA\ for Au. The mass densities used are based on the values reported by experimentalists as will be discussed posteriorly. Noble metal elements are similar in other aspects and are therefore considered closely related species that occupy the same column in the Periodic Table; due to this similarity it is assumed that studying them, and their alloys, can be predicted.

The approach we follow is based on our previous studies of amorphous materials where we succeeded in creating representative atomic structures. We use a supercell that contains enough atoms (108 silver atoms were considered initially) to make it as representative of the bulk as possible, within the limitations imposed by the computing resources needed and by the first-principles approach used. To create the alloys, we maintain the silver FCC structure and substitute silver atoms with gold atoms in a random fashion. By random we mean a personal selection of the sites hoping that no bias is introduced by this procedure. 

Once the samples were created (we constructed four samples for a given concentration) their structure was adjusted to the corresponding experimental density and was optimized to find the atomic positions that minimize the energy of the system. Then, the ``bands" were calculated using the Materials Studio suite of codes; in particular, the code used was DMol$^{3}$ that we have employed in our previous studies and that gives good results. We write ``bands" since due to the disordered structure of the alloys, periodicity is non-existent and therefore care should be exercised when dealing with crystalline concepts. Now we are ready to calculate the electronic Density of States (eDoS) for each atomic structure constructed, (remembering that for a given concentration more than one structure was constructed) and analyze the results to elucidate how dependent the eDoS is on the detailed distribution of solutes.

It is well known that, at sufficiently low temperatures, the constant-volume, $C_{v}$, and the constant-pressure, $C_{p}$, specific heats of metals are very similar \cite{Ashcroft1976}. The electronic contribution to the specific heat follows a linear behavior with temperature, $T$, at these low temperatures,
	\begin{equation}
		C_{v} = \gamma T,
		\label{eq:1}
	\end{equation}
where $\gamma$ is experimentally obtained from the above equation. Some approximate measured values for $\gamma$ in alkali metals are around 2 mJ mole$^{-1}$ K$^{-2}$, whereas for transition metals like manganese (Mn), $\gamma$ can be as high as 17 mJ mole$^{-1}$ K$^{-2}$ \cite{Ashcroft1976}.

The free-electron theory applied to metals provides an expression for $\gamma$ which only depends on the electron density of states at the Fermi level, $N(E_{F})$. In this approximation, the free-electron value, $\gamma_{f}$, is given by \cite{Sutton1993}:
	\begin{equation}
  		\gamma_{f} = \frac{\pi^2}{3} k_{B}^{2} N(E_{F}),
  		\label{eq:2}
	\end{equation}
where $k_{B}$ is the Boltzmann constant and the eDoS at the Fermi level, $N(E_{F})$ is proportional to the bare-electron mass, $m$, and to the electron number density, $n$, within a volume $V$ as follows \cite{Sutton1993, Cetina1977, Kaxiras2019}:
	\begin{equation}
		N(E_{F}) = \frac{V}{2\pi^2} \left( \frac{2m}{\hbar^2} \right)^{3/2} E_{F}^{1/2} = 3n \frac{m}{\hbar^2} \left( 3 \pi^2 n \right)^{-2/3}.
		\label{eq:3}
	\end{equation}

Since electrons in a metal are not exactly free, but are affected by the electron-phonon and electron-impurity interactions, it is not surprising that measured values for $\gamma$ generally are different from $\gamma_{f}$. To include these effects, it is customary to define a renormalized electron mass, $m^{*}$, such that the ratio $m^{*}/m$ corresponds to the ratio of the measured $\gamma$ to the calculated free-electron $\gamma_{f}$ \cite{Ashcroft1976, Ziman1960}:
	\begin{equation}
		\frac{m^{*}}{m} = \frac{\gamma}{\gamma_{f}}.
		\label{eq:4}
	\end{equation}

For some alkali metals, for which the free-electron theory is well suited, the ratio of equation \eqref{eq:4} is close to 1. For other metals the ratio can be as low as 1/20 (for bismuth (Bi)) or as high as 27 (for Mn), approximately \cite{Ashcroft1976}. For noble metals, early estimations for this ratio are 1.02 for pure Ag and 1.16 for pure Au \cite{Ziman1960, Keesom1956}, and no results for their alloys as a function of concentration have been reported.

\section{\label{sec:METHOD}METHOD}

Forty-two 108-atom supercells with an initial FCC structure were employed to study the electronic structure of the substitutionally disordered alloy Ag$_{100-x}$Au$_{x}$. Four different supercells were constructed for each concentration $x = 4, 13, 20, 30, 40, 50, 60, 70, 75, 90$ at.\%, by randomly substituting silver atoms with gold atoms within the crystalline FCC positions. For $x = 0, 100$, corresponding to the pure elements, only one 108-atom FCC supercell was employed. By modifying the lengths of the supercells' edges, while maintaining constant both the fractional atomic positions of solvents and solutes, and preserving the cubic nature of the structures, the volumes of the forty-two supercells were adjusted to reproduce the experimental densities \cite{Kraut2000} (see Table \ref{tab:table1}). Then, geometry optimization (GO) processes were performed on each supercell to ensure that the atoms located themselves in a minimum-energy position. These processes altered the locations of the atoms creating thereby a \textit{disordered} sample.
	\begin{table}[h]
	\caption{\label{tab:table1} Supercell densities for Ag$_{100-x}$Au$_{x}$. Taken from Ref. \cite{Kraut2000}.
	}%
		\begin{ruledtabular}
			\begin{tabular}{cccc}
				&  $x$ \textrm{{(}at. \%{)}} &
			\textrm{Density {(}g cm$^{-3}${)}} & \\
			\colrule
				& 0  & 10.5052 & \\
				& 4  & 10.8236 & \\
				& 13 & 11.6331 & \\
				& 20 & 12.2838 & \\
				& 30 & 13.1034 & \\
				& 40 & 14.0110 & \\
				& 50 & 14.9227 & \\
				& 60 & 15.8160 & \\
				& 70 & 16.7039 & \\
				& 75 & 17.1446 & \\
				& 90 & 18.4487 & \\
				& 100 & 19.2869 & \\
			\end{tabular}
		\end{ruledtabular}
	\end{table}

GO calculations were performed using the first-principles code DMol$^{3}$ \cite{Delley1995} included in the Dassault Systèmes BIOVIA Materials Studio software \cite{BioviaMS}. The convergence criteria for the GO processes were: $2.7 \times 10^{-4}$ eV for the energy, $5.4 \times 10^{-2}$ eV/\AA\ for the forces, and $5 \times 10^{-3}$ \AA\ for the maximum value of the displacements.

Single-point energies were calculated for each optimized supercell. The calculated eigenvalues for the electronic energies were smeared using Gaussians functions with a width of 0.2 eV and the electron Density of States, $N(E)$, was then obtained. Finally, from the value of $N(E)$ at the Fermi level, $\gamma_{f}$ was calculated through equation \eqref{eq:2}.

All calculations were spin-unrestricted and employed a $5 \times 5 \times 5$ k-point grid using a double numerical plus d-polarization (DND) basis. Core electrons were treated with the semi-core pseudopotentials (DSPP) included in the DMol$^{3}$ library and a real-space cutoff of 6.0 \AA\ was chosen for the electronic atomic functions. The exchange-correlation functional was on the Local Density Approximation (LDA) level of the theory according to Vosko, Wilk, and Nussair \cite{Vosko1980}. For the self-consistent-field calculations, the density convergence was set to a value of $1\times 10^{-6}$, and a thermal smearing to a value of 136 meV was applied.

\section{\label{sec:RESULTS}RESULTS}

\begin{figure}[h]
		\includegraphics{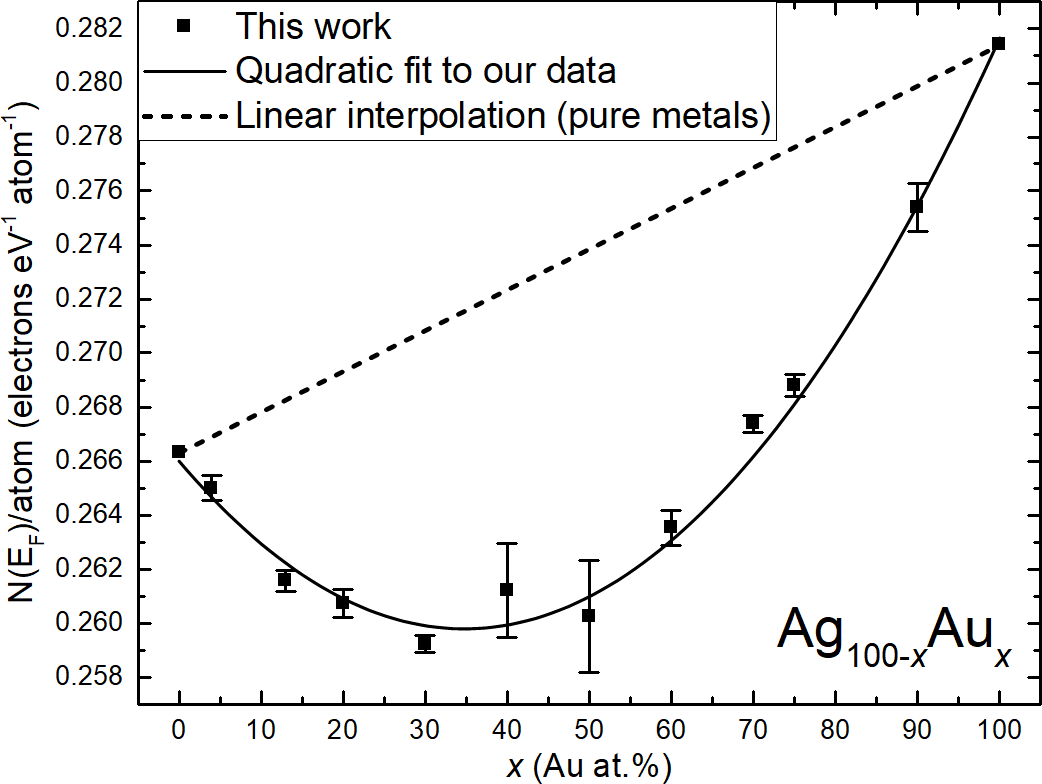}
		\caption{\label{fig:2} Electron Density of States at the Fermi level, $N(E_{F})$, for the Ag$_{100-x}$Au$_{x}$ alloys for the twelve atomic concentrations studied, $x$. The error bars represent the standard deviation for each concentration due to our averaging the four different supercells for a given concentration (see Table \ref{tab:table2}). The quadratic fit to our data is shown as the solid line and has an $R^{2}$ value of 0.9898. The dashed line is the linear interpolation between the pure metals.
}%
	\end{figure}

The $N(E_{F})$ was obtained for all forty-two supercells and then the values for each concentration were averaged (see Table \ref{tab:table2}). The values are shown in Figure \ref{fig:2} along with the quadratic adjustment $N(E_{F}) = A_{2} x^{2} + A_{1} x + A_{0}$, where $A_{2}=5.0246 \times 10^{-6}$, $A_{1}=-3.4337 \times 10^{-4}$, and $A_{0}=0.2659$, and with an $R^{2}$ value of 0.9898.  The quadratic fit to our data is shown as the solid line while the dashed line indicates the linear interpolation between the pure metals. The error bars included represent the standard deviation of the 4 calculated eDoS for each concentration.
	\begin{table}[h]
	\caption{\label{tab:table2}%
Average $N(E_{F})$, $\gamma_{f}$, and $\gamma_{p}$ for the Ag$_{100-x}$Au$_{x}$ alloys. The values for the pure constituents ($x=0$ and $x=100$) do not have standard deviations (See Figure 2) since only one 108-atom supercell was employed. $\gamma_{f}$ was calculated using equation \eqref{eq:2}. $\gamma_{p}$ was calculated using equations \eqref{eq:5} and \eqref{eq:6}.
	}%
		\begin{ruledtabular}
			\begin{tabular}{c c c c}
			$x$ & $N(E_{F})$ \textrm{per atom} & $\gamma_{f}$ & $\gamma_{p}$ \\
			\multicolumn{1}{l}{{(}at. \%{)}} & {(}electrons eV$^{-1}$ atom$^{-1}${)} & \multicolumn{2}{l}{{(}mJ mole$^{-1}$ K$^{-2}${)}} \\
			\colrule
			0  & 0.2663 $\pm\ \left(0.00 \times 10^{-4}\right)$ & 0.6276 & 0.6461 \\
			4  & 0.2650 $\pm\ \left(4.71 \times 10^{-4}\right)$ & 0.6245 & 0.6439 \\
			13 & 0.2616 $\pm\ \left(3.85 \times 10^{-4}\right)$ & 0.6165 & 0.6378 \\
			20 & 0.2607 $\pm\ \left(5.08 \times 10^{-4}\right)$ & 0.6144 & 0.6373 \\
			30 & 0.2592 $\pm\ \left(3.15 \times 10^{-4}\right)$ & 0.6108 & 0.6359 \\
			40 & 0.2612 $\pm\ \left(1.74 \times 10^{-3}\right)$ & 0.6156 & 0.6433 \\
			50 & 0.2602 $\pm\ \left(2.07 \times 10^{-3}\right)$ & 0.6132 & 0.6432 \\
			60 & 0.2635 $\pm\ \left(6.51 \times 10^{-4}\right)$ & 0.6210 & 0.6538 \\
			70 & 0.2674 $\pm\ \left(3.23 \times 10^{-4}\right)$ & 0.6302 & 0.6659 \\
			75 & 0.2688 $\pm\ \left(4.06 \times 10^{-4}\right)$ & 0.6335 & 0.6706 \\
			90 & 0.2754 $\pm\ \left(9.00 \times 10^{-4}\right)$ & 0.6490 & 0.6908 \\
			100 & 0.2814 $\pm\ \left(0.00 \times 10^{-4}\right)$ & 0.6632 & 0.7085 \\
			\end{tabular}
		\end{ruledtabular}
	\end{table}
	
To indirectly estimate $m^{*}$ by relating our $\gamma_{f}$ values with the experimental ones, we define $\gamma_{p}(x)$ as: 
	\begin{equation}
		\gamma_{p}(x) = F(x) \gamma_{f}(x),
		\label{eq:5}
	\end{equation}
where $F(x)$ is a linear function of $x$ such that $F(0)$ would yield $\gamma_{p}(0)$ to be equal to the experimental $\gamma$ for pure Ag, and $F(100)$ would yield $\gamma_{p}(100)$ to be equal to the experimental $\gamma$ for pure Au. Using our values we arrive at the following expression for $F(x)$:
	\begin{equation}
		F(x) = 1.0295 + \left( 3.88 \times 10^{-4} \right) x.
		\label{eq:6}
	\end{equation}
	
Values for $\gamma_{p}(x)$ for the other ten concentrations studied, calculated using $F(x)$ of equation \eqref{eq:6}, are shown in the last column of Table \ref{tab:table2}.

As shown in Figure \ref{fig:1}, the Au-rich part of the alloy was not studied in the work of Green and Valladares \cite{Green1966}. However, in 1966 following their methodology and using their equipment, Will and Green \cite{Will1966} performed studies on the specific heat of AuSn alloys and pure Au. Therefore, we include the $\gamma$ value for pure Au of Will and Green \cite{Will1966} into the Green and Valladares data to facilitate better comparisons with other experimental works that dealt with a broader range of concentrations. 

\begin{figure}[h]
		\includegraphics{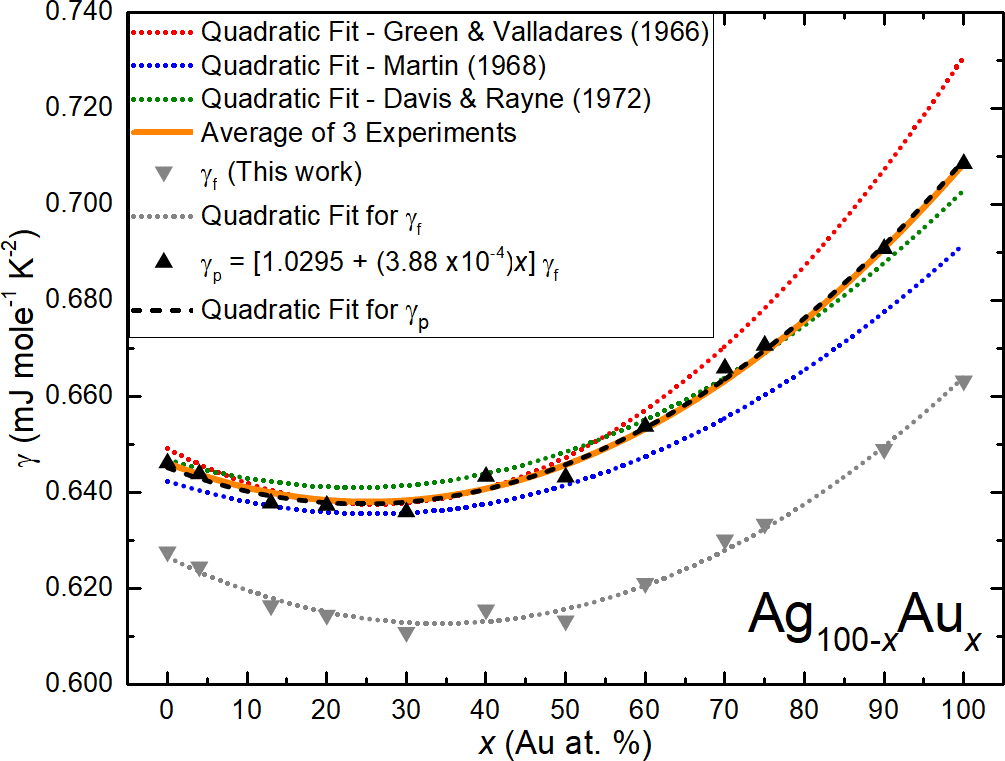}
		\caption{\label{fig:3} Experimental $\gamma$ values from three different groups. Dotted red line, extended Green and Valladares; dotted blue line, due to Martin; dotted green line, results by Davis and Rayne. The average results for the three experiments are shown by the orange solid line that agrees very nicely with our renormalized electron mass results, black triangles. The gray triangles represent our simulation results for the bare electron mass. Color online only.
}%
	\end{figure}

With this modification, the experimental results of Martin \cite{Martin1968}, Davis and Rayne \cite{Davis1972}, and the extended Green and Valladares curve are shown in Figure \ref{fig:3}. In this figure, quadratic fits to the experiments of Green and Valladares \cite{Green1966} (dotted red line), Martin \cite{Martin1968} (dotted blue line), and Davis and Rayne \cite{Davis1972} (dotted green line) are included. The value for $\gamma$ of pure gold in the Green and Valladares line was taken from Will and Green \cite{Will1966}. The average line for the three experiments is shown as the solid orange line. Gray triangles represent the values for $\gamma_{f}$ obtained from our calculated $N(E_{F})$ (see Table \ref{tab:table2}). The gray dotted line is shown as a guide to the eye. Black triangles are our calculated values of $\gamma_{p}$ according to equation \eqref{eq:5}. See Table \ref{tab:table3} for the parameters of the quadratic fits.

	\begin{table}[h]
	\caption{\label{tab:table3} Quadratic fit parameters to $\gamma = B_{2} x^{2} + B_{1} x + B_{0}$.
	}%
		\begin{ruledtabular}
			\begin{tabular}{lcccc}
			 &
			$B_{2} \times 10^{-5}$ &
			$B_{1} \times 10^{-4}$ &
			$B_{0}$ &
			$R^{2}$ \\
			\colrule
			Green \& Valladares \cite{Green1966} & 1.7066 & -8.8998 & 0.6491 & 0.9735 \\
			Martin \cite{Martin1968}             & 1.0204 & -5.2667 & 0.6424 & 0.9873 \\
			Davis \& Rayne \cite{Davis1972}      & 1.0588 & -4.9786 & 0.6469 & 0.9887 \\
			Average of experiments               & 1.2619 & -6.3817 & 0.6461 & 1 \\
			Fit for $\gamma_{p}$                 & 1.2553 & -6.1489 & 0.6451 & 0.9952 \\
			\end{tabular}
		\end{ruledtabular}
	\end{table}

\begin{figure}[h]
		\includegraphics{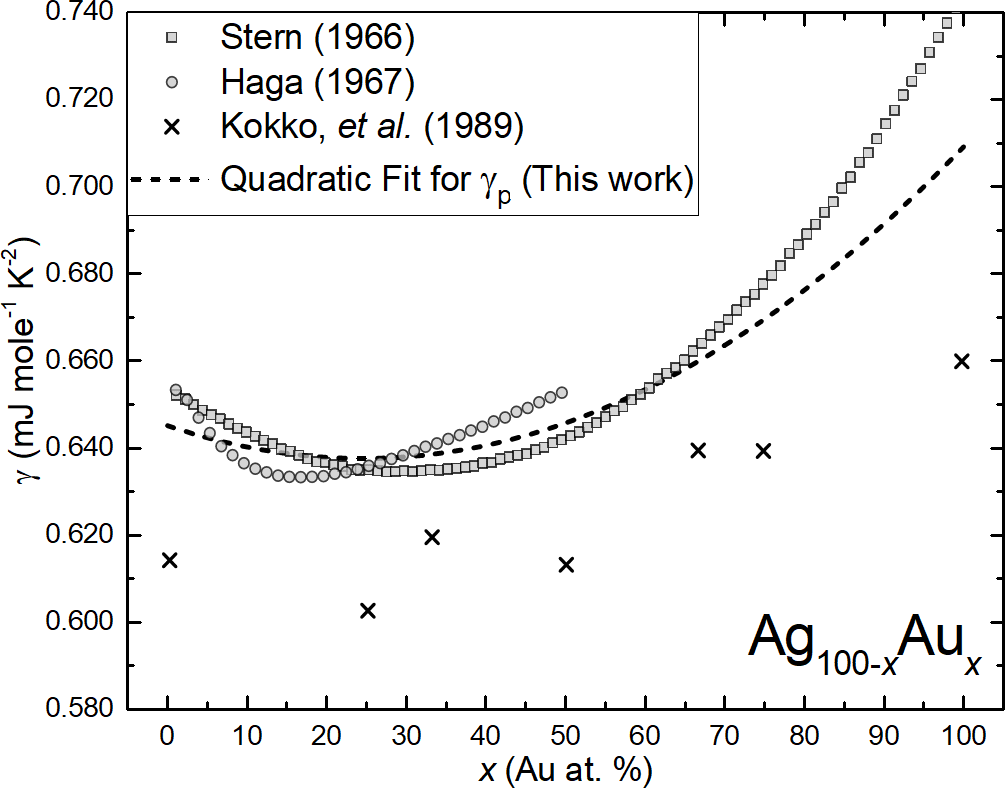}
		\caption{\label{fig:4} Theoretical $\gamma$ values from Stern \cite{Stern1966} (gray squares), Haga \cite{Haga1967} (gray circles), and Kokko \textit{et al.} \cite{Kokko1989} (crosses). Our result for $\gamma_{p}$ from equation \eqref{eq:5} is also shown (dashed black line).
}%
	\end{figure}

Figure \ref{fig:4} displays the theoretical results reported by Stern \cite{Stern1966} who invoked the charging hypothesis around solutes and solvents, the calculations reported by Haga \cite{Haga1967} arguing that the experimental results are due to the renormalization of the electron mass caused by the electron-phonon interaction, and the results by Kokko \textit{et al.} \cite{Kokko1989} who used crystalline structures and claimed that as a first approximation their LMTO calculations may represent what occurs in these alloys, ``at least qualitatively". Our results for $\gamma_{p}$ from equation \eqref{eq:5} are also shown as the dashed black line.

\begin{figure}[h]
		\includegraphics{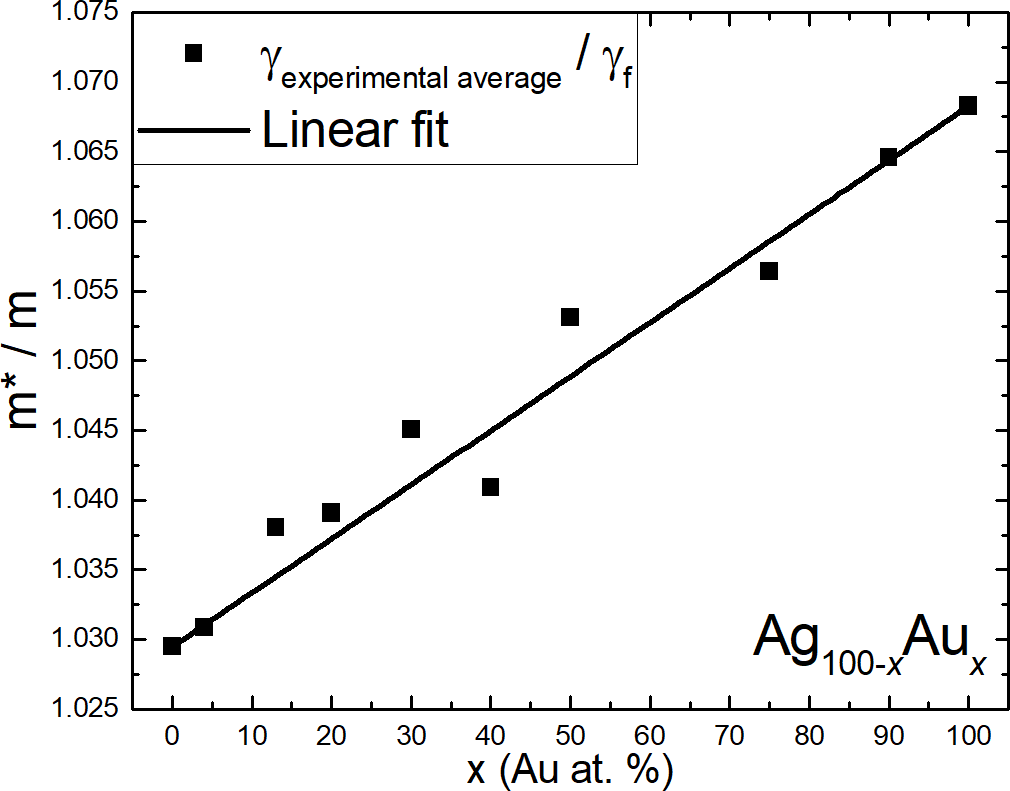}
		\caption{\label{fig:5} Ratio of the renormalized electron mass to free electron mass, $m^{*}/m$, obtained by dividing the experimental average of $\gamma$ (see Table \ref{tab:table3} and Fig. \ref{fig:2}) by our free value $\gamma_{f}$ (Table \ref{tab:table2}) The linear fit has the equation $m^{*}/m = 1.0295 + \left(3.88  \times 10^{-4}\right) x $. Values for $m^{*}/m$ for pure Ag and pure Au are: 1.02 and 1.16, respectively \cite{Ziman1960, Keesom1956}. 
}%
	\end{figure}

A byproduct of our calculations is an estimate of the intensity of the electron-phonon interaction. Although our computational results display the parabolic behavior observed in the experimental results, they do not exactly coincide with them. There is the necessity to include a multiplicative factor of order 1 that varies with concentration, and based on some isolated results for the pure elements it is of the order of magnitude of the renormalization of the electron-mass. This factor is shown in Figure \ref{fig:5} and varies linearly between the pure-elements values. This is the first time an indirect calculation of the non-bare electron mass is reported.

\section{\label{sec:CONCLUSIONS}CONCLUSIONS}
\textit{Ab initio} calculations of the electronic structure were carried out for forty-two 108-atom supercells of the disordered structures of the binary alloy Ag$_{100-x}$Au$_{x}$ as a function of $x$. Since four samples were constructed for each ``impure" concentration, and since the resulting structures are different, it was necessary to investigate what the effect of this difference had on the eDoS of the impure samples. The calculations for the pure samples were only done for one structure. It was found that the four different, initially stochastic, distribution of solvent and solute gave results for the eDoS very similar, as shown by the error bars displayed in Figure \ref{fig:2}, bars that represent the standard deviation for each concentration due to our averaging of the results for the four cells. The largest dispersion occurs for concentrations in the vicinity of 50\% which indicates that the tendency to order ...Ag-Au-Ag-Au... that may occur at this concentration is seriously affecting the disorder, creating perhaps ordered clusters within the material.

Our calculations are highly representative of the real phenomenon, since it starts with a large FCC supercell of Ag, which is initially crystalline, and then gold atoms are randomly substituted. The density is adjusted to match experimental results, and then a geometry optimization process is performed to foster the final structure. These supercells are now as close to the real ones as needed to represent the disordered alloy, they are not crystalline, no Bloch functions are used, the disordered is introduced by randomly substituting the solute atoms and by the geometry optimization runs. That is why we obtain the parabolic behavior found experimentally.%

As a byproduct of these calculations we find that in order to obtain an excellent agreement with experiment it is necessary to consider the mass renormalization of the electron displayed in Figure \ref{fig:5} where the behavior indicates that the order of magnitude of the values coincide with some reported experimentally. This is a first since, to our knowledge, no estimation of the renormalization has been obtained before for these binary metallic systems.

Although all theoretical results do manifest the curvature displayed by the experiment, it is clear that the physical reasons invoked are different and therefore it becomes difficult to ascribe the observed behavior to one of them; also, the fits are not as good as desired. Our results unequivocally display this behavior and show that they are independent of the mechanism involved, although to find a good fit a factor assumed to be due to the electron-phonon interaction has been included giving, indirectly, good results and information concerning the behavior of this renormalization phenomenon.

The procedure that we have displayed above indicates that old problems can be revisited with the support of new tools to understand a variety of phenomena that have not been fully comprehended. What is new? The old is new.

\begin{acknowledgments}
D.H.-R. thanks DGAPA-UNAM for his postdoctoral fellowship. I.R. thanks CONAHCyT for his postdoctoral fellowship. This work was supported by UNAM Postdoctoral Program (POSDOC). A.A.V., R.M.V., and A.V. thank DGAPA-UNAM (PAPIIT) for continued financial support to carry out research projects under Grants No. IN116520 and IN118223. M. T. Vázquez and O. Jiménez provided the information requested. A. Pompa assisted with the technical support and maintenance of the computing unit at IIM-UNAM. Simulations were partially performed at the Computing Center of DGTIC-UNAM under the project LANCAD-UNAM-DGTIC-131. 
\end{acknowledgments}

\bibliography{AgAuAlloys_Revisited}

\begin{thebibliography}{20}%
\makeatletter
\providecommand \@ifxundefined [1]{%
 \@ifx{#1\undefined}
}%
\providecommand \@ifnum [1]{%
 \ifnum #1\expandafter \@firstoftwo
 \else \expandafter \@secondoftwo
 \fi
}%
\providecommand \@ifx [1]{%
 \ifx #1\expandafter \@firstoftwo
 \else \expandafter \@secondoftwo
 \fi
}%
\providecommand \natexlab [1]{#1}%
\providecommand \enquote  [1]{``#1''}%
\providecommand \bibnamefont  [1]{#1}%
\providecommand \bibfnamefont [1]{#1}%
\providecommand \citenamefont [1]{#1}%
\providecommand \href@noop [0]{\@secondoftwo}%
\providecommand \href [0]{\begingroup \@sanitize@url \@href}%
\providecommand \@href[1]{\@@startlink{#1}\@@href}%
\providecommand \@@href[1]{\endgroup#1\@@endlink}%
\providecommand \@sanitize@url [0]{\catcode `\\12\catcode `\$12\catcode
  `\&12\catcode `\#12\catcode `\^12\catcode `\_12\catcode `\%12\relax}%
\providecommand \@@startlink[1]{}%
\providecommand \@@endlink[0]{}%
\providecommand \url  [0]{\begingroup\@sanitize@url \@url }%
\providecommand \@url [1]{\endgroup\@href {#1}{\urlprefix }}%
\providecommand \urlprefix  [0]{URL }%
\providecommand \Eprint [0]{\href }%
\providecommand \doibase [0]{https://doi.org/}%
\providecommand \selectlanguage [0]{\@gobble}%
\providecommand \bibinfo  [0]{\@secondoftwo}%
\providecommand \bibfield  [0]{\@secondoftwo}%
\providecommand \translation [1]{[#1]}%
\providecommand \BibitemOpen [0]{}%
\providecommand \bibitemStop [0]{}%
\providecommand \bibitemNoStop [0]{.\EOS\space}%
\providecommand \EOS [0]{\spacefactor3000\relax}%
\providecommand \BibitemShut  [1]{\csname bibitem#1\endcsname}%
\let\auto@bib@innerbib\@empty
\bibitem [{\citenamefont {Valladares}\ and\ \citenamefont
  {Green~Jr.}(1965)}]{Valladares1965}%
  \BibitemOpen
  \bibfield  {author} {\bibinfo {author} {\bibfnamefont {A.~A.}\ \bibnamefont
  {Valladares}}\ and\ \bibinfo {author} {\bibfnamefont {B.~A.}\ \bibnamefont
  {Green~Jr.}},\ }\bibfield  {title} {\bibinfo {title} {{KG14}. {E}lectronic
  {S}pecific {H}eat of {S}ilver-{G}old {A}lloys},\ }\href@noop {} {\bibfield
  {journal} {\bibinfo  {journal} {Bull. Am. Phys. Soc.}\ }\textbf {\bibinfo
  {volume} {10}},\ \bibinfo {pages} {127} (\bibinfo {year} {1965})}\BibitemShut
  {NoStop}%
\bibitem [{\citenamefont {Green~Jr.}\ and\ \citenamefont
  {Valladares}(1966)}]{Green1966}%
  \BibitemOpen
  \bibfield  {author} {\bibinfo {author} {\bibfnamefont {B.~A.}\ \bibnamefont
  {Green~Jr.}}\ and\ \bibinfo {author} {\bibfnamefont {A.~A.}\ \bibnamefont
  {Valladares}},\ }\bibfield  {title} {\bibinfo {title} {{L}ow-{T}emperature
  {S}pecific {H}eats of {AgAu} {A}lloys},\ }\href
  {https://doi.org/10.1103/PhysRev.142.379} {\bibfield  {journal} {\bibinfo
  {journal} {Phys. Rev.}\ }\textbf {\bibinfo {volume} {142}},\ \bibinfo {pages}
  {379} (\bibinfo {year} {1966})}\BibitemShut {NoStop}%
\bibitem [{\citenamefont {Stern}(1966)}]{Stern1966}%
  \BibitemOpen
  \bibfield  {author} {\bibinfo {author} {\bibfnamefont {E.~A.}\ \bibnamefont
  {Stern}},\ }\bibfield  {title} {\bibinfo {title} {{C}harging and the
  {P}roperties of {A}lloys},\ }\href {https://doi.org/10.1103/PhysRev.144.545}
  {\bibfield  {journal} {\bibinfo  {journal} {Phys. Rev.}\ }\textbf {\bibinfo
  {volume} {144}},\ \bibinfo {pages} {545} (\bibinfo {year}
  {1966})}\BibitemShut {NoStop}%
\bibitem [{\citenamefont {Friedel}(1954)}]{Friedel1954}%
  \BibitemOpen
  \bibfield  {author} {\bibinfo {author} {\bibfnamefont {J.}~\bibnamefont
  {Friedel}},\ }\bibfield  {title} {\bibinfo {title} {Electronic structure of
  primary solid solutions in metals},\ }\href
  {https://doi.org/10.1080/00018735400101233} {\bibfield  {journal} {\bibinfo
  {journal} {Adv. Phys.}\ }\textbf {\bibinfo {volume} {3}},\ \bibinfo {pages}
  {446} (\bibinfo {year} {1954})}\BibitemShut {NoStop}%
\bibitem [{\citenamefont {Haga}(1967)}]{Haga1967}%
  \BibitemOpen
  \bibfield  {author} {\bibinfo {author} {\bibfnamefont {E.}~\bibnamefont
  {Haga}},\ }\bibfield  {title} {\bibinfo {title} {Electronic specific heats of
  silver-gold alloys},\ }\href {https://doi.org/10.1088/0370-1328/91/1/325}
  {\bibfield  {journal} {\bibinfo  {journal} {Proc. Phys. Soc.}\ }\textbf
  {\bibinfo {volume} {91}},\ \bibinfo {pages} {169} (\bibinfo {year}
  {1967})}\BibitemShut {NoStop}%
\bibitem [{\citenamefont {Kokko}\ \emph {et~al.}(1989)\citenamefont {Kokko},
  \citenamefont {Ojala},\ and\ \citenamefont {Mansikka}}]{Kokko1989}%
  \BibitemOpen
  \bibfield  {author} {\bibinfo {author} {\bibfnamefont {K.}~\bibnamefont
  {Kokko}}, \bibinfo {author} {\bibfnamefont {E.}~\bibnamefont {Ojala}},\ and\
  \bibinfo {author} {\bibfnamefont {K.}~\bibnamefont {Mansikka}},\ }\bibfield
  {title} {\bibinfo {title} {{F}ermi {L}evel {D}ensity of {S}tates in {A}g-{A}u
  {A}lloys},\ }\href {https://doi.org/10.1002/pssb.2221530124} {\bibfield
  {journal} {\bibinfo  {journal} {Phys. Status Solidi B}\ }\textbf {\bibinfo
  {volume} {153}},\ \bibinfo {pages} {235} (\bibinfo {year}
  {1989})}\BibitemShut {NoStop}%
\bibitem [{\citenamefont {Clark}\ and\ \citenamefont
  {Dawber}(1972)}]{Clark1972}%
  \BibitemOpen
  \bibfield  {author} {\bibinfo {author} {\bibfnamefont {J.~A.}\ \bibnamefont
  {Clark}}\ and\ \bibinfo {author} {\bibfnamefont {P.~G.}\ \bibnamefont
  {Dawber}},\ }\bibfield  {title} {\bibinfo {title} {Theory of alloys, a
  coherent pseudopotential model},\ }\href
  {https://doi.org/10.1088/0305-4608/2/5/017} {\bibfield  {journal} {\bibinfo
  {journal} {J. Phys. F: Met. Phys.}\ }\textbf {\bibinfo {volume} {2}},\
  \bibinfo {pages} {930} (\bibinfo {year} {1972})}\BibitemShut {NoStop}%
\bibitem [{\citenamefont {Martin}(1968)}]{Martin1968}%
  \BibitemOpen
  \bibfield  {author} {\bibinfo {author} {\bibfnamefont {D.~L.}\ \bibnamefont
  {Martin}},\ }\bibfield  {title} {\bibinfo {title} {{S}pecific {H}eat below
  3°{K} of {S}ilver-{G}old {A}lloys},\ }\href
  {https://doi.org/10.1103/PhysRev.176.790} {\bibfield  {journal} {\bibinfo
  {journal} {Phys. Rev.}\ }\textbf {\bibinfo {volume} {176}},\ \bibinfo {pages}
  {790} (\bibinfo {year} {1968})}\BibitemShut {NoStop}%
\bibitem [{\citenamefont {Davis}\ and\ \citenamefont
  {Rayne}(1972)}]{Davis1972}%
  \BibitemOpen
  \bibfield  {author} {\bibinfo {author} {\bibfnamefont {T.~H.}\ \bibnamefont
  {Davis}}\ and\ \bibinfo {author} {\bibfnamefont {J.~A.}\ \bibnamefont
  {Rayne}},\ }\bibfield  {title} {\bibinfo {title} {Specific {H}eat and
  {R}esidual {R}esistivity of {B}inary and {T}ernary {N}oble-{M}etal
  {A}lloys},\ }\href {https://doi.org/10.1103/PhysRevB.6.2931} {\bibfield
  {journal} {\bibinfo  {journal} {Phys. Rev. B}\ }\textbf {\bibinfo {volume}
  {6}},\ \bibinfo {pages} {2931} (\bibinfo {year} {1972})}\BibitemShut
  {NoStop}%
\bibitem [{\citenamefont {Ashcroft}\ and\ \citenamefont
  {Mermin}(1976)}]{Ashcroft1976}%
  \BibitemOpen
  \bibfield  {author} {\bibinfo {author} {\bibfnamefont {N.~W.}\ \bibnamefont
  {Ashcroft}}\ and\ \bibinfo {author} {\bibfnamefont {N.~D.}\ \bibnamefont
  {Mermin}},\ }\href@noop {} {\emph {\bibinfo {title} {{Solid State
  Physics}}}}\ (\bibinfo  {publisher} {Brooks/Cole. Cengage Learning},\
  \bibinfo {address} {Belmont, CA, USA},\ \bibinfo {year} {1976})\ pp.\
  \bibinfo {pages} {48,49}\BibitemShut {NoStop}%
\bibitem [{\citenamefont {Sutton}(1993)}]{Sutton1993}%
  \BibitemOpen
  \bibfield  {author} {\bibinfo {author} {\bibfnamefont {A.~P.}\ \bibnamefont
  {Sutton}},\ }\href@noop {} {\emph {\bibinfo {title} {{Electronic Structure of
  Materials}}}}\ (\bibinfo  {publisher} {Oxford University Press},\ \bibinfo
  {address} {New York, USA},\ \bibinfo {year} {1993})\ pp.\ \bibinfo {pages}
  {160--162}\BibitemShut {NoStop}%
\bibitem [{\citenamefont {Cetina}\ \emph {et~al.}(1977)\citenamefont {Cetina},
  \citenamefont {Magaña},\ and\ \citenamefont {Valladares}}]{Cetina1977}%
  \BibitemOpen
  \bibfield  {author} {\bibinfo {author} {\bibfnamefont {E.}~\bibnamefont
  {Cetina}}, \bibinfo {author} {\bibfnamefont {F.}~\bibnamefont {Magaña}},\
  and\ \bibinfo {author} {\bibfnamefont {A.~A.}\ \bibnamefont {Valladares}},\
  }\bibfield  {title} {\bibinfo {title} {The free-electron gas in $n$
  dimensions},\ }\href {https://doi.org/10.1119/1.10859} {\bibfield  {journal}
  {\bibinfo  {journal} {Am. J. Phys.}\ }\textbf {\bibinfo {volume} {45}},\
  \bibinfo {pages} {960} (\bibinfo {year} {1977})}\BibitemShut {NoStop}%
\bibitem [{\citenamefont {Kaxiras}\ and\ \citenamefont
  {Joannopoulos}(2019)}]{Kaxiras2019}%
  \BibitemOpen
  \bibfield  {author} {\bibinfo {author} {\bibfnamefont {E.}~\bibnamefont
  {Kaxiras}}\ and\ \bibinfo {author} {\bibfnamefont {J.~D.}\ \bibnamefont
  {Joannopoulos}},\ }\href@noop {} {\emph {\bibinfo {title} {{Quantum Theory of
  Materials}}}}\ (\bibinfo  {publisher} {Cambridge University Press},\ \bibinfo
  {address} {Cambridge, UK},\ \bibinfo {year} {2019})\ pp.\ \bibinfo {pages}
  {110,111}\BibitemShut {NoStop}%
\bibitem [{\citenamefont {Ziman}(1960)}]{Ziman1960}%
  \BibitemOpen
  \bibfield  {author} {\bibinfo {author} {\bibfnamefont {J.~M.}\ \bibnamefont
  {Ziman}},\ }\href@noop {} {\emph {\bibinfo {title} {{Electrons and Phonons:
  The Theory of Transport Phenomena in Solids}}}}\ (\bibinfo  {publisher}
  {Oxford University Press},\ \bibinfo {address} {New York, USA},\ \bibinfo
  {year} {1960})\ pp.\ \bibinfo {pages} {112--114}\BibitemShut {NoStop}%
\bibitem [{\citenamefont {Keesom}\ and\ \citenamefont
  {Pearlman}(1956)}]{Keesom1956}%
  \BibitemOpen
  \bibfield  {author} {\bibinfo {author} {\bibfnamefont {P.~H.}\ \bibnamefont
  {Keesom}}\ and\ \bibinfo {author} {\bibfnamefont {N.}~\bibnamefont
  {Pearlman}},\ }\bibinfo {title} {Low {T}emperature {H}eat {C}apacity of
  {S}olids},\ in\ \href@noop {} {\emph {\bibinfo {booktitle} {Low Temperature
  Physics I / K{\"a}ltephysik I}}},\ \bibinfo {editor} {edited by\ \bibinfo
  {editor} {\bibfnamefont {S.}~\bibnamefont {Fl{\"u}gge}}}\ (\bibinfo
  {publisher} {Springer Berlin Heidelberg},\ \bibinfo {address} {Berlin,
  Heidelberg},\ \bibinfo {year} {1956})\ pp.\ \bibinfo {pages}
  {282--337}\BibitemShut {NoStop}%
\bibitem [{\citenamefont {Kraut}\ and\ \citenamefont
  {Stern}(2000)}]{Kraut2000}%
  \BibitemOpen
  \bibfield  {author} {\bibinfo {author} {\bibfnamefont {J.~C.}\ \bibnamefont
  {Kraut}}\ and\ \bibinfo {author} {\bibfnamefont {W.~B.}\ \bibnamefont
  {Stern}},\ }\bibfield  {title} {\bibinfo {title} {The density of
  gold-silver-copper alloys and its calculation from the chemical
  composition},\ }\href {https://doi.org/10.1007/BF03216580} {\bibfield
  {journal} {\bibinfo  {journal} {Gold Bull.}\ }\textbf {\bibinfo {volume}
  {33}},\ \bibinfo {pages} {52} (\bibinfo {year} {2000})}\BibitemShut {NoStop}%
\bibitem [{\citenamefont {Delley}(1995)}]{Delley1995}%
  \BibitemOpen
  \bibfield  {author} {\bibinfo {author} {\bibfnamefont {B.}~\bibnamefont
  {Delley}},\ }\bibfield  {title} {\bibinfo {title} {{DM}ol, a standard tool
  for density functional calculations: {R}eview and advances},\ }in\ \href@noop
  {} {\emph {\bibinfo {booktitle} {Modern Density Functional Theory}}},\
  Vol.~\bibinfo {volume} {2},\ \bibinfo {editor} {edited by\ \bibinfo {editor}
  {\bibfnamefont {J.~M.}\ \bibnamefont {Seminario}}\ and\ \bibinfo {editor}
  {\bibfnamefont {P.}~\bibnamefont {Politzer}}}\ (\bibinfo  {publisher}
  {Elsevier},\ \bibinfo {year} {1995})\ pp.\ \bibinfo {pages}
  {221--254}\BibitemShut {NoStop}%
\bibitem [{Bio(2015)}]{BioviaMS}%
  \BibitemOpen
  \href@noop {} {\bibinfo {title} {{Dassault Systèmes}. {BIOVIA} {M}aterials
  {S}tudio ({R}elease 2016-1)}} (\bibinfo {year} {2015})\BibitemShut {NoStop}%
\bibitem [{\citenamefont {Vosko}\ \emph {et~al.}(1980)\citenamefont {Vosko},
  \citenamefont {Wilk},\ and\ \citenamefont {Nusair}}]{Vosko1980}%
  \BibitemOpen
  \bibfield  {author} {\bibinfo {author} {\bibfnamefont {S.~H.}\ \bibnamefont
  {Vosko}}, \bibinfo {author} {\bibfnamefont {L.}~\bibnamefont {Wilk}},\ and\
  \bibinfo {author} {\bibfnamefont {M.}~\bibnamefont {Nusair}},\ }\bibfield
  {title} {\bibinfo {title} {Accurate spin-dependent electron liquid
  correlation energies for local spin density calculations: a critical
  analysis},\ }\href {https://doi.org/10.1139/p80-159} {\bibfield  {journal}
  {\bibinfo  {journal} {Can. J. Phys.}\ }\textbf {\bibinfo {volume} {58}},\
  \bibinfo {pages} {1200} (\bibinfo {year} {1980})}\BibitemShut {NoStop}%
\bibitem [{\citenamefont {Will}\ and\ \citenamefont {Green}(1966)}]{Will1966}%
  \BibitemOpen
  \bibfield  {author} {\bibinfo {author} {\bibfnamefont {T.~A.}\ \bibnamefont
  {Will}}\ and\ \bibinfo {author} {\bibfnamefont {B.~A.}\ \bibnamefont
  {Green}},\ }\bibfield  {title} {\bibinfo {title} {Specific {H}eats of {A}u
  and {AuSn} at {L}ow {T}emperatures},\ }\href
  {https://doi.org/10.1103/PhysRev.150.519} {\bibfield  {journal} {\bibinfo
  {journal} {Phys. Rev.}\ }\textbf {\bibinfo {volume} {150}},\ \bibinfo {pages}
  {519} (\bibinfo {year} {1966})}\BibitemShut {NoStop}%
\end{thebibliography}%

\end{document}